\documentclass[12pt]{iopart}

\usepackage{bm}
\usepackage{graphicx}
\usepackage{color}
\usepackage{subfigure}
\usepackage{hyperref}
\usepackage{latexsym}
\usepackage{amsthm}
\usepackage{amssymb}
\usepackage{braket}
\usepackage{cite}
\DeclareGraphicsExtensions{.jpg,.pdf, .mps, .png, .eps, .ps, .EPS,.gif}

\begin{document}

\def\be{\begin{equation}}
\def\ee{\end{equation}}

\def\bc{\begin{center}}
\def\ec{\end{center}}
\def\bea{\begin{eqnarray}}
\def\eea{\end{eqnarray}}
\newcommand{\avg}[1]{\langle{#1}\rangle}
\newcommand{\Avg}[1]{\left\langle{#1}\right\rangle}

\def\ie{\textit{i.e.}}
\def\etal{\textit{et al.}}
\def\m{\vec{m}}
\def\G{\mathcal{G}}

\newcommand{\davide}[1]{{\bf\color{blue}#1}}
\newcommand{\gin}[1]{{\bf\color{green}#1}}
\title[Simplicial complexes: higher-order spectral dimension  and dynamics]{{Simplicial complexes:} higher-order spectral dimension  and dynamics }

\author{Joaqu\'in J. Torres}
\address{Department of Electromagnetism and Physics of the Matter and Institute Carlos I for Theoretical and Computational Physics, University of Granada, E-18071 Granada, Spain}

\author{Ginestra Bianconi}

\address{School of Mathematical Sciences, Queen Mary University of London, London, E1 4NS, United Kingdom\\
Alan Turing Institute, The British Library, London, United Kingdom}
\ead{ginestra.bianconi@gmail.com}
\vspace{10pt}
\begin{indented}
\item[]
\end{indented}

\begin{abstract}
Simplicial complexes constitute the underlying  topology of interacting complex systems including among the others brain and social interaction networks. They are generalized network structures that allow to go beyond the framework of pairwise interactions and to capture the many-body interactions between two or more nodes strongly affecting dynamical processes.  In fact, the simplicial complexes topology allows to assign a dynamical variable not only to the nodes of the interacting complex systems but also to links, triangles, and so on. Here we show evidence that the dynamics defined on simplices of different dimensions can be significantly different even if we compare dynamics of simplices belonging to the same simplicial complex.  By investigating the spectral properties of the simplicial complex model called ``Network Geometry with Flavor'' we provide evidence that the up and down higher-order Laplacians can have a finite spectral dimension whose value increases  as the  order of the Laplacian increases. Finally we  discuss the implications of this result for higher-order diffusion  defined on simplicial complexes. 
\end{abstract}

%
%
%
%
%

\section{Introduction}
Simplicial complexes are generalized network structures that allow to capture the many body interactions existing between the constituents of complex systems \cite{Perspective,Lambiotte,Bassett}.
They are becoming increasingly popular to represent brain data \cite{Bassett,BlueBrain,Petri}, social interacting systems  \cite{Barrat,Latora,Kahng_epidemics,Arenas_epidemics}, financial networks \cite{Aste1,Aste2} and complex materials \cite{Bassett_granular,Tadic}, beyond the framework of pairwise interactions.
 A simplicial complex is formed by a set of simplices such as nodes, links, triangles, tetrahedra and so on glued to each other along their faces. Being built by geometrical building blocks, simplicial complexes represent an ideal setting to investigate the properties of emergent network geometry and topology in complex systems \cite{Perspective,Emergent,NGF,Hyperbolic}.Moreover they reveal the rich interplay between network geometry and dynamics\cite{Ana1,Ana2,Ana3,Arenas1,Arenas2,Bick,Kuehn}.
 
The recently proposed non-equilibrium growing simplicial complex model called ``Network Geometry with Flavor'' (NGF) \cite{NGF} is able to display emergent hyperbolic network geometry \cite{Hyperbolic} together with the major universal properties of complex networks including scale-free degree distribution, small-word distance property, high clustering coefficient and significant modular structure.
 Interestingly, the simplicial complexes generated by the NGF model display also a finite spectral dimension \cite{Ana1,Ana2,Polytopes,Ren}.
 The spectral dimension \cite{Burioni1,Burioni2,Burioni3,Kahng} characterises the spectrum of the graph Laplacian of network geometries and is well known to affect the return-time probability of classical \cite{Burioni1} critical phenomena \cite{Ising,Erzan} and quantum diffusion \cite{Blumen}. Additionally  the spectral dimension  strongly affects {the synchronization properties} of the Kuramoto model which display a thermodynamically stable synchronized phase only if the spectral dimension $d_S$ is greater than four \cite{Ana1,Ana2}. Finally the spectral dimension is also used in quantum gravity to probe the geometry of different model of quantum space-time \cite{Loll1,Loll2,dario1,dario2,Jonsson1,Jonsson2}

Recent works  \cite{Ziff,Ana3,Barbarossa,simplices2} have emphasised that simplicial complexes can sustain dynamical processes whose variables can be located not only on their nodes but also on their higher dimensional simplices {such as} links, triangles and so on.
In particular, in Ref. \cite{Ana3} the Kuramoto model has been extended to treat synchronization of phases located in higher-dimensional simplices. Additionally, a higher-order diffusion dynamics has been defined over simplicial complexes \cite{simplices2}. 
The higher-order diffusion dynamics and the higher-order Kuramoto model depend on the higher-order boundary maps of the simplices and the higher-order Laplacian matrix.
The higher-order Laplacian matrix  \cite{Barbarossa,simplices2, thesis,Jost} of order $n>0$ describes a diffusion dynamics taking place between simplices of order $n$ and can be decomposed in the sum between the up-Laplacian and the down-Laplacian.
The  higher-order discrete Laplacian has been studied by several mathematicians  \cite{thesis,Jost} , however as far as we know, there is no previous result showing that the high-order Laplacian can display a finite spectral dimension.

In this work we investigate the spectral properties of the higher-order Laplacian on NGF. We show that the higher-order up and down-Laplacians have a finite spectral dimension that increases with their order $n$.
By investigating the properties of higher-order diffusion on NGF we find that the higher-order spectral dimension has a significant effect on the return-time probability of the process. Therefore, we provide evidence that the diffusion, occurring on the same simplicial complex but taking place on simplices of different {order} $n,$ can {induce  significantly} different dynamical behavior.
\section{Methods}

\subsection{Simplicial complexes }

 Simplicial complexes are able to capture higher-order interactions between two or more nodes described by simplices. A $n$-dimensional {\it simplex} $\mu$ is formed by $n+1$ nodes
\bea
\mu=[i_0,i_1,\dots,i_{n}].
\eea 
Therefore, a $0$-dimensional simplex is a node, a $1$-dimensional simplex is a link, and so on.
A face of $n$-dimensional simplex is a $n^{\prime}$-dimensional simplex formed by a proper subset of $n^{\prime}+1$ nodes of the original simplex. {Consequently,} we necessarily have $n^{\prime}<n$.
A simplicial complex ${\mathcal K}$ is   a set of simplices that is closed under the inclusion of the faces of the simplices.
We will indicate with $d$ the dimension of the simplicial complex determining the maximum dimension of its simplices.
Moreover, we will indicate with $N_{[n]}$ the number of $n$-dimensional simplices of the simplicial complex ${\mathcal K}$. Therefore in the following $N_{[0]}$ indicates the number of nodes, $N_{[1]}$  the number of links, $N_{[2]}$  the number of triangles and so on.
Simplicial complexes can sustain a diffusion dynamics occurring on its $n$-dimensional faces. This higher-order diffusion dynamics is determined by the properties of the higher-order Laplacians. In order to introduce here the higher-order Laplacian we will devote the next paragraph to some fundamental quantities in network topology.
\subsection{Oriented simplices and boundary map}

In topology each  $n$-dimensional simplex $\mu$ 
\bea
\mu=[i_0,i_1,\dots,i_{n}].
\eea 
has an orientation given by the sign of the permutation of the label of the nodes. Therefore, we have  
\bea
[i_0,i_1,\ldots, i_n]=(-1)^{\sigma(\pi)}[i_{\pi(0)},i_{\pi(1)},\dots,i_{\pi(n)}]
\eea
where $\sigma(\pi)$ indicates the parity of the permutation ${\bf \pi}$.

The boundary map $\partial_n$ is a linear operator acting on linear combinations of  $n$-dimensional {simplices and defined by  its action on each of the simplices $\mu=[i_0,i_2\ldots,i_n]$ of the simplicial complex as}
\bea
\partial_n [i_0,i_1\ldots,i_n]=\sum_{p=0}^n(-1)^p[i_0,i_1,\dots,i_{p-1},i_{p+1},\dots,i_n].
\eea
Therefore the boundary map of a link $[i,j]$ is given by 
\bea
\partial_1 [i,j]=[j]-[i].
\eea
Similarly the boundary map of a triangle $[i,j,k]$ is given by 
\bea
\partial_2[i,j,k]=[j,k]-[i,k]+[i,j].
\eea
From this definition it follows directly that 
\bea
\partial_{n-1}\partial_n={\bf 0},
\label{boundary}
\eea
relations that is often referred to by saying that the "boundary of the boundary is zero".
For instance, the reader can easily check using the above definitions that  $\partial_{1} \partial_2 [i,j,k]=0$.
The boundary map $\partial_n$ {can also be}  represented by the incidence matrix ${\bf B}_{[n]}$ of dimension $N_{[n-1]}\times N_{[n]}$. {Then,} Eq.(\ref{boundary}) can be expressed using the incidence matrices  as
\bea
{\bf B}_{[n-1]}{\bf B}_{[n]}={\bf 0}.
\label{boundary2}
\eea
\begin{figure}[h!]
\centering
\includegraphics[width=8cm]{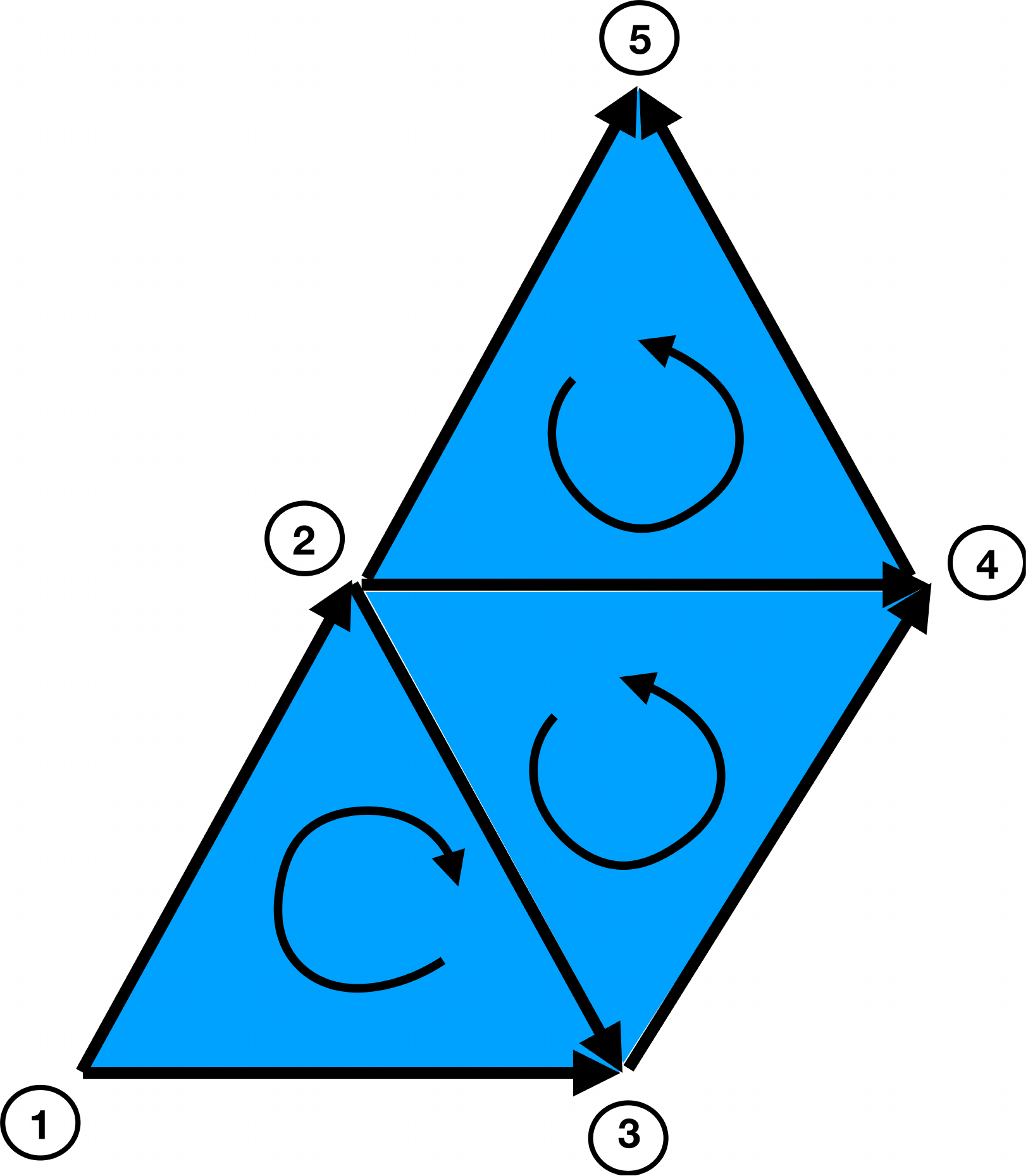}
	\caption{We show a small {$2$-dimensional} simplicial complex of $N_{[0]}=5$ nodes, $N_{[1]}=7$ links and $N_{[3]}=3$ triangles whose incidence matrices ${\bf B}_{[1]}$ and ${\bf B}_{[2]}$ are given in Eqs. (\ref{b1}) and (\ref{b2}).}
	\label{Figure1}
	\end{figure}
	
In Figure $\ref{Figure1}$ we show {a small $2$-dimensional} simplicial complex formed by the set of nodes $\{[1],[2],[3],[4], [5]\}$, the set of links $\{[12],[13],[23],[34],[24], [25],[45]\}$ and triangles $\{[123],[234],[245]\}$.
Its boundary maps are given by 

\bea
{\bf B}_{[1]}=\left(\begin{array}{ccccccc}
-1&-1 &0&0&0&0&0\\
1&0&-1&0&-1&-1&0\\
0&1&1&-1&0&0&0\\
0&0&0&1&1&0&-1\\
0&0&0&0&0&1&1\\
\end{array}\right),
\label{b1}
\eea
\bea
{\bf B}_{[2]}=\left(\begin{array}{ccc}
1&0&0\\
-1&0&0\\
1&1&0\\
0&1&0\\
0&-1&1\\
0&0&-1\\
0&0&1\\
\end{array}\right).
\label{b2}
\eea

\subsection{Higher-order Laplacians}

The zero-order Laplacian   ${\bf L}_{[0]}$ is the usual graph Laplacian defined on networks and is a $N_{[0]}\times N_{[0]}$ matrix of elements
\bea
\left(L_{[0]}\right)_{ij}=\delta_{ij}-a_{ij},
\eea
where here and in the following $\delta_{ij}$ indicates the Kronecker delta, and $a_{ij}$ indicates the element $(i,j)$ of the adjacency matrix.
The graph Laplacian  ${\bf L}_{[0]}$ can be also  expressed in terms of the incidence matrix ${\bf B}_{[1]}$ as 
\bea
{\bf L}_{[0]}={\bf B}_{[1]}{\bf B}^{\top}_{[1]}.
\eea 
This definition can be extended to higher-order Laplacians using higher-order incidence matrices ${\bf B}_{[n]}$.
In particular the higher-order Laplacian  ${\bf L}_{[n]}$ (with $n>0$) \cite{simplices2,thesis,Barbarossa} is the  $N_{[n]}\times N_{[n]}$ matrix defined as 
\bea
{\bf L}_{[n]}={\bf L}^{down}_{[n]}+{\bf L}^{up}_{[n]},
\eea
with 
\bea
{\bf L}^{down}_{[n]}&=&{\bf B}^{\top}_{[n]}{\bf B}_{[n]},\nonumber \\
{\bf L}^{up}_{[n]}&=&{\bf B}_{[n+1]}{\bf B}^{\top}_{[n+1]}.
\eea
The higher-order Laplacians are independent on the orientation of the simplices as long as the orientation of the simplices is induced by the label of the nodes.
The degeneracy of the zero eigenvalue of the  $n$ Laplacian ${\bf L}_{[n]}$ is equal to the Betti number $\beta_n$. The eigenvectors associated to the zero eigenvalue of the $n$-Laplacian  are localized on  the corresponding $n$-dimensional cavities of the simplicial complex. {Therefore, the higher-order Laplacians with $n>0$ are not guaranteed to have a zero eigenvalue as simplicial complexes with  $\beta_n=0$ for some $n>0$ exist.} In particular, if the topology of the simplicial complex is trivial, i.e. $\beta_0=1$ and $\beta_n=0$ for all $n>0,$ the  Laplacians of order $n>0$ do not admit a zero eigenvalue.

Another important property of the $n$-Laplacian is that each non-zero eigenvalue is either a non-zero eigenvalue of the $n$-order up-Laplacian or is a non-zero eigenvalue of the $n$-order down-Laplacian.
Consider an eigenvector ${\bf v}$ of the up-Laplacian with eigenvalue $\lambda\neq 0.$ {Then, we have} 
\bea
{\bf B}_{[n+1]}{\bf B}_{[n+1]}^{\top}{\bf v}=\lambda {\bf v},
\eea 
or equivalently 
\bea
 {\bf v}=\frac{1}{\lambda}{\bf B}_{[n+1]}{\bf B}_{[n+1]}^{\top}{\bf v}.
\eea
Let us apply the down-Laplacian to the eigenvector ${\bf v}.$ {Thus
we obtain}
\bea
{\bf B}^{\top}_{[n]}{\bf B}_{[n]}{\bf v}=\frac{1}{\lambda}{\bf B}^{\top}_{[n]}{\bf B}_{[n]}{\bf B}_{[n+1]}{\bf B}_{[n+1]}^{\top}{\bf v}={\bf 0},
\eea
where we have used Eq. (\ref{boundary2}).
It follows that if ${\bf v}$ is an eigenvector associated to a non-zero eigenvalue $\lambda$ of the $n$-order up-Laplacian then it is an eigenvector of the $n$-order down-Laplacian with zero eigenvalue. Therefore,  in this case ${\bf v}$ is an eigenvector  of the $n$-order Laplacian with the eigenvalue $\lambda$.
Similarly, it can be easily shown that if ${\bf v}$ is an eigenvector associated to a non-zero eigenvalue $\lambda$ of the $n$-order down-Laplacian  then it is also an eigenvector  of the $n$-order Laplacian with the same eigenvalue.
Consequently the spectrum of the $n$-order Laplacian includes all the eigenvalues of the $n$-order up-Laplacian and the $n$-order down-Laplacian.

Another important property of the high-order up and down Laplacians is  that the spectrum of the $n$-order up-Laplacian coincides with the spectrum of the $(n+1)$-order down-Laplacian as the two are related by 
\bea
{\bf L}_{[n]}^{up}=\left({\bf L}^{down}_{[n+1]}\right)^{\top}.
\label{transpose}
\eea
The $n$-Laplacian is positive semi-definite  {and, therefore,} it has $N_{[n]}$ non negative eigenvalues that we indicate as 
\bea
0\leq \lambda_1\leq \lambda_2\leq \ldots \lambda_r\leq \ldots \leq \lambda_{N_{[n]}}.
\eea
Moreover, in the following we will  indicate by ${\bf v}^{(r)}$ the eigenvector corresponding to eigenvalue $\lambda_r$ of the $n$-Laplacian.

\subsection{Spectral dimension of the graph Laplacian}

The spectral dimension is defined for networks ($1$-dimensional simplicial complexes) with distinct geometrical properties, and determines the dimension of the network as ``experienced'' by a diffusion process taking place on it \cite{Ana1,Ana2,Blumen,Burioni1,Burioni2,Burioni3}.
The spectral dimension is traditionally defined starting from the 
 density of eigenvalues $\rho(\lambda)$ of the  $0$-Laplacian. We say that a network has {\it spectral dimension} $d_S^{[0]}$ if the density of eigenvalues  $\rho(\lambda)$ of the  $0$-Laplacian follows the scaling relation
\bea
\rho(\lambda)\simeq \tilde{C}_{[0]}\lambda^{d_S^{[0]}/2-1}
\label{scaling}
\eea
for $\lambda\ll1$.
In $d$-dimensional Euclidean lattices $d_S^{[0]}=d$. Additionally,  $d_S^{[0]}$ is related to  the Hausdorff dimension $d_H$ of the network by the inequalities \cite{Jonsson1,Jonsson2}
\bea
d_H\geq d_S^{[0]}\geq 2\frac{d_H}{d_H+1}. \label{eq:dHdS}
\eea
Therefore, for small-world networks, which have infinite Hausdorff dimension $d_H=\infty$, it is only possible to have finite spectral dimension $d_S^{[0]}\geq 2$.
If the density of eigenvalues $\rho(\lambda)$ follows Eq.(\ref{scaling}) it {results} that the cumulative distribution $\rho_c(\lambda)$ evaluating the density of eigenvalues $\lambda'\leq \lambda$ satisfies
\bea
\rho_c(\lambda)\simeq C_{[0]}\lambda^{d_S^{[0]}/2}, 
\label{eq:rho_c}
\eea
for $\lambda\ll 1$. In presence of a finite spectral dimension the Fiedler eigevalue $\lambda_2$ satisifies
\bea
\lambda_{2}\propto N_{[0]}^{-2/d_S^{[0]}}.
\label{Fidler}
\eea
Therefore, the Fidler eigenvalue $\lambda_2\to 0$ as $N_{[0]}\to \infty$ and we say that in the large network limit the spectral gap closes. This is another distinct property of networks with a geometrical character, i.e. significantly different from random graphs and expanders.\\
The spectral dimension has been proven to be essential to determine the stability of the synchronized state of the Kuramoto model   which can be thermodynamically stable only if
$
d_S^{[0]}>4.
$

In the next section we will show  that the  notion of spectral dimension can be generalized to  order $n>0$ with important consequences for higher-order simplicial complex dynamics.

\section{Results}

In this section we will investigate the spectral properties of a recently proposed model of simplicial complexes called ``Network Geometry with Flavor''.
We will show that the higher-order up-Laplacians of these simplicial complexes display a finite spectral dimension depending on the  {order $n$} of the up-Laplacian considered, the dimension of the simplicial complex $d$ and a parameter of the model called flavor $s$.
Therefore given a single instance of a NGF we can define different spectral dimensions $d_S^{[n]}$ for $0<n<d-1$. Here we will show that this implies that the dynamics defined on simplices of different dimension $n$ of the same simplicial complex can be significantly different.

\subsection{Network Geometry with Flavor}
The model ``Network Geometry with Flavor'' (NGF)   \cite{NGF,Hyperbolic}  generates $d$-dimensional simplicial complexes. Each simplex is obtained  by performing a non-equilibrium process consisting in the continuous addition of new  $d$-simplices attached to the rest of the simplicial complex along a single $(d-1)$-face.
To every $(d-1)$-face $\mu$ of the simplicial complex, (i.e. a link for $d=2$, or a triangular face for $d=3$) we associate an {\em incidence number} $n_{\mu}$ given by the number of $d$-dimensional simplices incident to it minus one. 
The  evolution of NGF  depends on a parameter $s\in\{-1,0,1\}$ called {\em flavor}. Starting from  a single $d$-dimensional  simplex, with $d\geq 2$ at each time we add a  $d$-dimensional simplex to a $(d-1)$-face $\mu$. The face $\mu$ is chosen randomly with   probability  $\Pi_{\mu}$ given by 
\bea
\Pi_{\mu}=\frac{1+s n_{\mu}}{\sum_{\nu}1+s n_{\nu}}.
\label{prob}
\eea

According to a classical result in combinatorics, under this dynamics we obtain a discrete manifold only if  $n_{\mu}$ can take exclusively the values $n_{\mu}=0,1$. 
This occurs only for $s=-1$.
In fact for $n_{\mu}=0$ we obtain  $\Pi_{\mu}={1}/{(\sum_{\nu}1+s n_{\nu})}>0$ but for $n_{\mu}=1$ we obtain $\Pi_{\mu}=0$. Therefore, the resulting simplicial complex is a discrete manifold, with each $(d-1)$-face incident at most  to two $d$-dimensional simplices, i.e. $n_{\mu}=0,1$. 
For $s=0$ the attachment probability $\Pi_{\mu}$ is uniform while for $s=1$ the attachment probability $\Pi_{\mu}$ increases linearly with the number of simplices already incident to the face $\mu$ implementing a generalized preferential attachment.
Therefore for  $s=0$ as for $s=1$ the incidence number $n_{\mu}$ can  take arbitrary large values $n_{\mu}=0,1,2,3\ldots$.

This model generates emergent hyperbolic geometry, and the underlying network is small-word (has infinite Hausdorff dimension, i.e. $d_H=\infty$), has high clustering coefficient and  high modularity. 
Interestingly, this model reduces to well known models in specific cases:
for $d=1$ and $s=1$ it reduces to the tree Barabasi-Albert model \cite{BA}, for $d=2$ and $s=0$ it reduces to the model first studied in Ref. \cite{Samukhin} and finally for $d=3$ and $s=-1$ it reduces to the random Apollonian model \cite{apollonian,apollonian2}.

\subsection{Spectral properties of NGF}

The  graph Laplacian of NGFs has been recently show to display a finite spectral dimension $d_S^{[0]}$ and localized eigenvectors with important consequences on dynamics \cite{Polytopes,Ana1,Ana2}.
Interestingly, here we show that also the higher-order up-Laplacians ${\bf L}_{[n]}^{up}$ and the higher-order down-Laplacians ${\bf L}_{[n]}^{down}$ of NGFs display a  finite spectral dimension.

\begin{figure}
\begin{center}
\includegraphics[width=15cm]{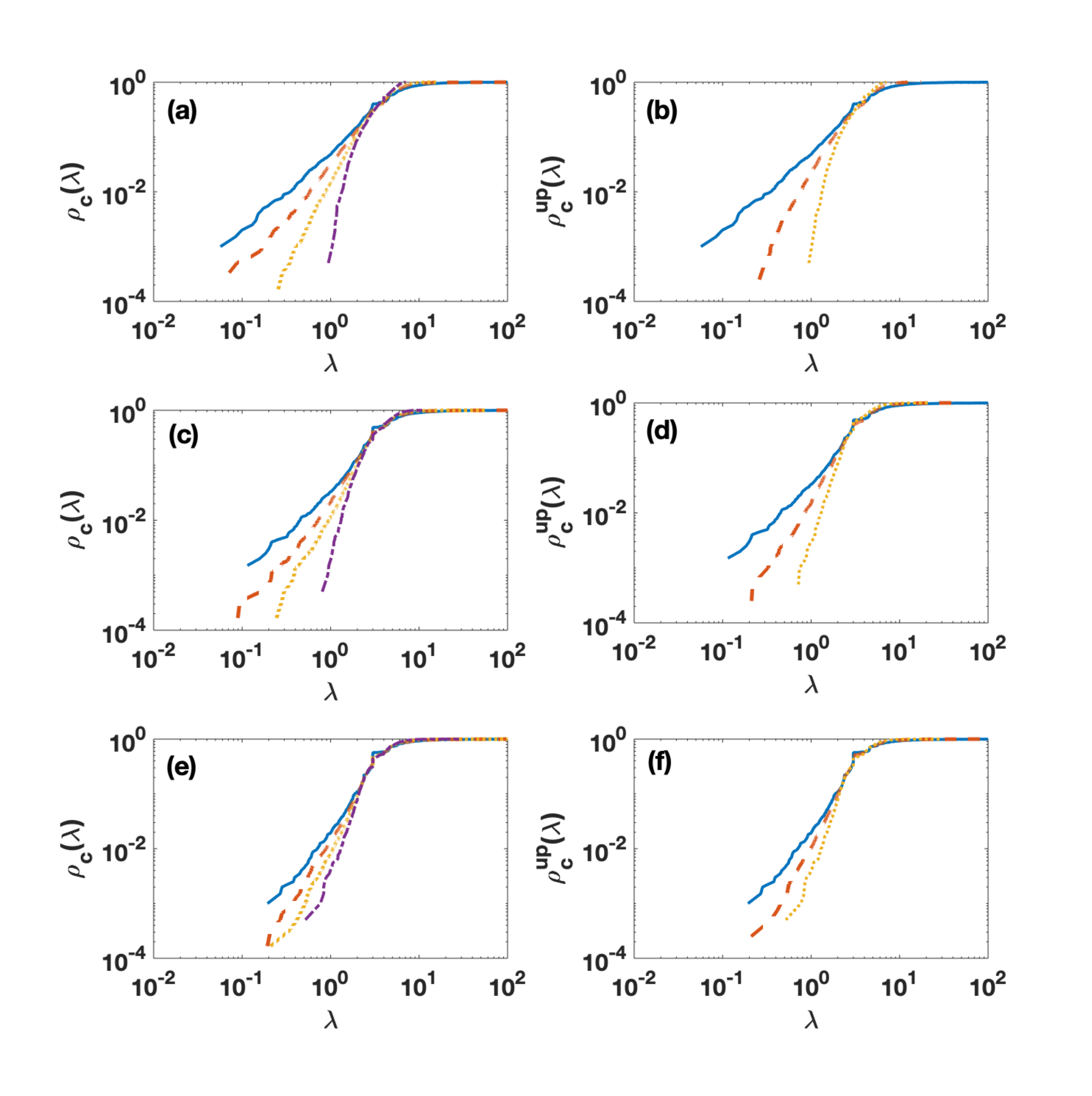}
\end{center}
\caption{  The cumulative density  of eigenvalues $\rho_c(\lambda)$ of the $n$-order Laplacian and the cumulative density  of non-zero eigenvalues $\rho_c^{up}(\lambda)$ of the $n$-order up-Laplacians are shown for the NGF with $N_{[0]}=2000$ nodes, $d=3$ and flavor $s=-1$ {(panels a and b), $s=0$ (panels c and d) and $s=1$ (panels e and f)}. Here the blue solid lines indicate $n=0$, the red  dashed lines indicate $n=1,$ the yellow dotted line indicates $n=2$ and the purple dot-dashed lines indicates $n=3$.}

\label{Higher_spectral_dimension}
\end{figure}

Since the up-Laplacian is defined as ${\bf L}^{up}_{[n]}={\bf B}_{[n+1]}{\bf B}^{\top}_{[n+1]}$ the eigenvalues $\lambda$ of the $n$-order up-Laplacian are the square of the singular values of the incidence matrix ${\bf B}_{[n+1]}$. The incidence matrix ${\bf B}_{[n+1]}$ is a rectangular $N_{[n]}\times N_{[n+1]}$ matrix, {therefore the non-zero singular values cannot be more than $\min(N_{[n]},N_{[n+1]})$.}  
For NGFs, that have a trivial topology, the Hodge decomposition \cite{Barbarossa} guarantees that the number ${\mathcal N}_{[n]}$ of non-zero eigenvalues of the $n$-order up-Laplacian with $n>0$ achieves this limit {and consequently we have}
 \bea
{\mathcal N}_{[n]}=\left\{\begin{array}{lcc}N_{[0]}-1& \mbox{if}& n=0,\\\min(N_{[n]},N_{[n+1]})& \mbox{if}& 0<n<d.\end{array}\right.
\eea
In Figure $\ref{Higher_spectral_dimension}$ we plot the cumulative density of eigenvalues $\rho_c(\lambda)$ of the $n$-order Laplacian and the cumulative density of non-zero eigenvalues $\rho_c^{up}(\lambda)$ of the $n$-order up-Laplacians of NGF with $d=3$ and flavor $s=-1,0,1$. The $n$-order up-Laplacians display a finite spectral dimension, i.e. their cumulative density of eigenvalues obeys the scaling 
\bea
\rho_c^{up}(\lambda)\simeq C_{[n]}\lambda^{d_S^{[n]}/2},
\eea
for $\lambda\ll1$.
The  fitted values of these higher-order spectral dimensions are  reported in Table $\ref{table}$ for different values of the order $n$ and the flavor $s$ of the {$3$-dimensional} NGF.
From this table it can be clearly shown that the values  of the higher-order spectral dimension $d_S^{[n]}$  increase with $n$
i.e.
\bea
d_S^{[n+1]}>d_S^{[n]}
\eea for any value of the flavor $s$ and have values greater than two. 
We note that our numerical results (not shown) clearly show that   this property remains valid also for NGFs of dimensions $d\neq 3$.

Since the $n$-order up-Laplacian is the transpose matrix of the $(n+1)$-order down-Laplacian ({as given in Eq. (\ref{transpose})}) the two matrices have the same spectrum. It follows directly that the $(n+1)$-order down-Laplacian has spectral dimension $d_S^{[n]}$.

From these results on the higher-order up-Laplacian we can easily determine the scaling of the density of eigenvalues for the higher-order Laplacian of NGFs.
In particular  for $0<n<d$ we have
\bea
\rho_c(\lambda)=\frac{{\mathcal N}_{[n-1]}}{N_{[n]}}C_{[n-1]}\lambda^{d_S^{[n-1]}/2}+\frac{{\mathcal N}_{[n]}}{N_{[n]}}C_{[n]}\lambda^{d_S^{[n]}/2},
\label{sc1}
\eea
for $n=0$ we have instead
\bea
\rho_c(\lambda)=C_{[0]}\lambda^{d_S^{[0]}/2},
\label{sc2}
\eea
and for $n=d$ we have  \bea
\rho_c(\lambda)=C_{[d-1]}\lambda^{d_S^{[d-1]}/2}.
\label{sc3}
\eea
Therefore the density of eigenvalues $\rho(\lambda)$ of the $n$-order Laplacian reads
\bea
\rho(\lambda)=\left\{\begin{array}{lcc}
\tilde{C}_{[0]}\lambda^{d_S^{[0]}/2-1} & \mbox{for} &n=0,\\
\tilde{C}_{[n-1]}\lambda^{d_S^{[n-1]}/2-1} +\tilde{C}_{[n]}\lambda^{d_S^{[n]}/2-1} & \mbox{for} &0<n<d,\\
 \tilde{C}_{[d-1]}\lambda^{d_S^{[d-1]}/2-1} & \mbox{for} & n=d,\\
\label{scrho}
\end{array}\right.
\eea
where $\tilde{C}_{[n]}$ are constants.
Therefore, the cumulative density of the eigenvalues of the higher-order Laplacian will asymptotically scale as a power-law dictated by the minimum between $d_S^{[n-1]}$ and $d_S^{[n]}$.

\begin{figure}
\begin{center}
\includegraphics[width=15cm]{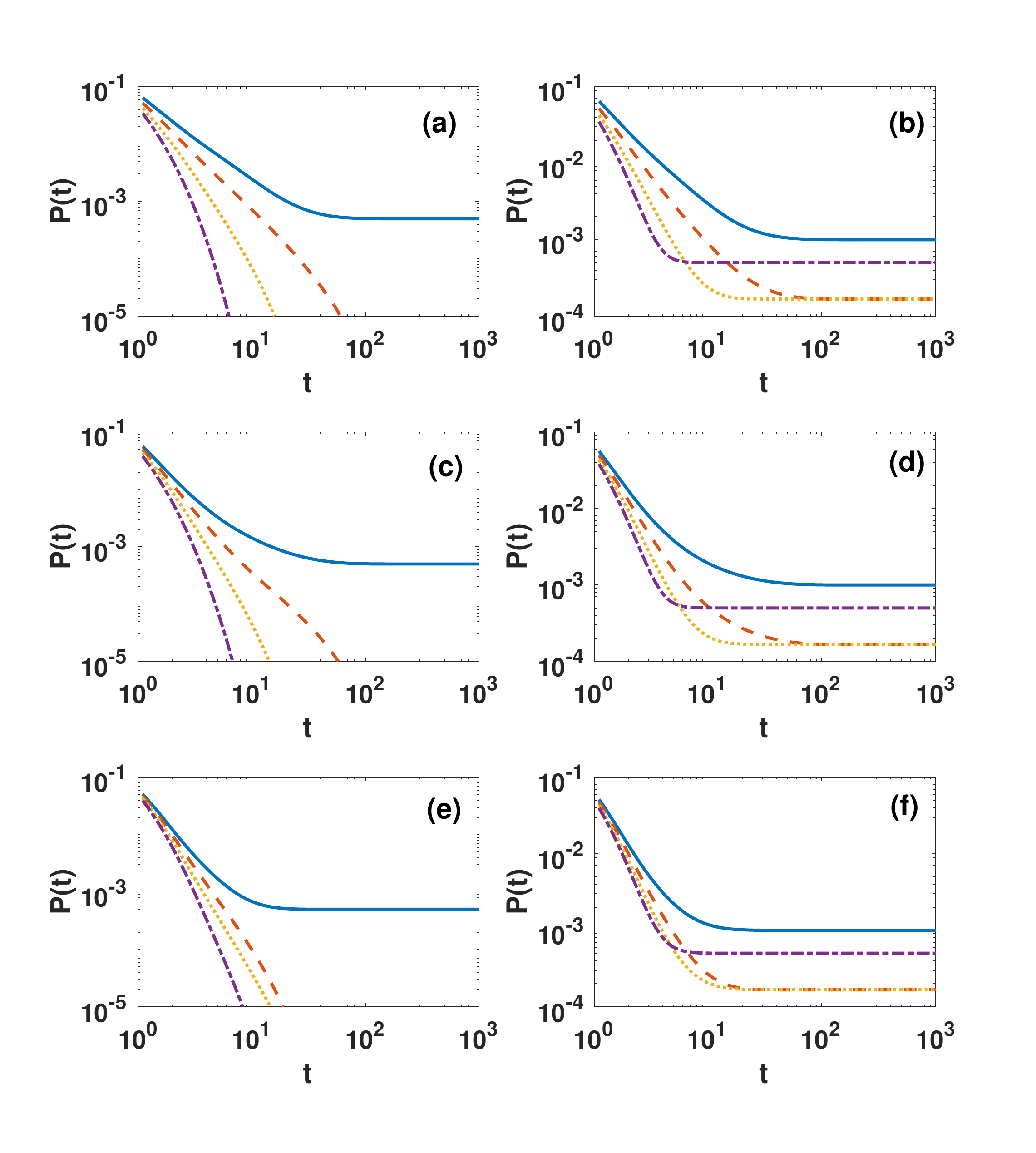}
\end{center}
\caption{The return-time probability $P(t)$ of the higher-order diffusion dynamics determined by Eq. (\ref{dif1}) (panels a, c, e) and Eq.(\ref{dif2}) (panels b, d, f) is shown for NGF with $N_{[0]}=2000$ nodes,  $d=3$ and flavor $s=-1$ {(panels a and b), $s=0$ (panels c and d) and $s=1$ (panels e and f)}. Here the blue solid lines indicate $n=0$, the red  dashed lines indicate $n=1,$ the yellow dotted line indicates $n=2$ and the purple dot-dashed line indicates $n=3.$}

\label{Return:fig}
\end{figure} 

\begin{table}
\caption{\label{table} Fitted value of the spectral dimension $d_S^{[n]}$ of the $n$-order up-Laplacian of the NGF for different values of $n$ and $s$ and  for $d=3$. The fitted values have been estimated from a single realization of the NGF with $N_{[0]}=2000$ nodes. The error over the fitted spectral dimension is the $0.01$ confidence interval of the corresponding linear regression model.}
\begin{tabular*}{\textwidth}{@{}l*{15}{@{\extracolsep{0pt plus
12pt}}l}}
\br
$d$/$s$&$n=0$&$n=1$&$n=2$\\
\mr
$s=-1$&$3.03\pm 0.03$&$5.36\pm0.04$ &$14.3\pm 0.5$\\
$s=0$&$3.57\pm 0.05$&$5.48\pm0.02$&$11.3\pm0.03$\\
$s=1$&$4.82\pm 0.08$&$6.04\pm0.05$&$7.8\pm 0.3$\\
\br
\end{tabular*}
\end{table}

\subsection{Diffusion using higher-order Laplacian}
Higher-order Laplacians ${\bf L}_{[n]}$ can be used to define a diffusion process defined over $n$-dimensional simplices.
For instance, one can consider a classical quantity $x_{\mu}$ defined on the $n$-dimensional simplices $\mu$ of the simplicial complex and use the $n$-Laplacian ${\bf L}_{[n]}$ to study its diffusion using the dynamical equation 
\bea
\frac{dx_{\mu}}{dt}=-\sum_{\nu\in S_{n}}\left({\bf L}_{[n]}\right)_{\mu,\nu}x_{\nu}.
\label{dif1}
\eea
where   with $S_{n}$ we indicate the set of all simplices of dimension $n$ (of cardinality $|S_{n}|=N_{[n]}$). For $n=0$ there is always a stationary state as $\beta_1$  indicates at the same time the number of connected components of the simplicial complex (therefore we have $\beta_1\geq 1$) and the degeneracy of the zero eigenvalue of the Laplacian matrix ${\bf L}_{[0]}$.  Additionally, for a connected network the stationary state is uniform over all the nodes of the network. However, Eq. (\ref{dif1}) for $n>0$ will have a stationary state only if the Betti number $\beta_n>0$, i.e. if there is at least a $n$-dimensional cavity in the simplicial complex. Note, however, that also if this stationary state exists  the stationary state will be non-uniform over the network but localized on the $n$-dimensional cavities. In order to describe a diffusion equation that has a non trivial stationary state also when $\beta_n=0$ we can modify the diffusion equation and consider instead the dynamics 
\bea
\frac{dx_{\mu}}{dt}=-\sum_{\nu\in S_n}\left({\bf L}_{[n]}\right)_{\mu,\nu}x_{\nu}-\lambda_1{ v}^{(1)}_\mu\sum_{\nu\in S_n}{ v}^{(1)}_{\nu}x_{\nu}.
\label{dif2}
\eea
This equation reduces to Eq. (\ref{dif1}) if the smallest eigenvalue of the $n$-Laplacian is zero (i.e. $\lambda_1=0$) and  admits always a non-trivial stationary state localized along the eigenvector ${\bf v}^{(1)}$ corresponding to the smallest eigenvalue.
The NGFs have Betti numbers $\beta_0=1$ and $\beta_n=0$ for every $n>0$. In this case, when $n>0$ the dynamics defined by Eq. (\ref{dif1}) gives a transient to a vanishing solution $x_{\nu}=0$ for every $n$-dimensional face $\nu$. On the contrary, the dynamics defined by Eq. (\ref{dif2}) gives a transient to a non-vanishing steady state solution.
{The solution for the two dynamical equations (\ref{dif1}) and (\ref{dif2}) can be written as}
\bea
x_{\mu}(t)=\sum_{r=1}^{N_{[n]}}e^{-\lambda_r (1-c\delta_{r,1}) t}v_\mu^{(r)}\sum_{\nu\in S_n}v_{\nu}^{(r)}x_{\nu}(0)
\eea 
where for the dynamics defined in Eq. (\ref{dif1}) we put $c=0$ while for the dynamics defined in Eq.(\ref{dif2}) we put $c=1$.

 For both dynamics, we investigate  the return-time probability $P(t)$ as a function of time. The return-time probability $P(t)$ is defined as the probability that the diffusion process  starting from a localized configuration on a  given simplex $\mu$ returns back to the simplex $\mu$ at time $t$, averaged over all simplices $\mu\in S_n$ of the simplicial complex.
 Therefore $P(t)$ is given by
 \bea
 P(t)&=&\sum_{\mu\in S_n}\sum_{r=1}^{N_{[n]}}e^{-\lambda_r (1-c\delta_{r,1}) t}v_\mu^{(r)}v_{\mu}^{(r)}=\sum_{r=1}^{N_{[n]}}e^{-\lambda_r (1-c\delta_{r,1}) t}
 \eea
 where in the last expression we have used the normalization of the eigenvectors ${\bf v}^{(r)},$ i.e.
 \bea
 \sum_{\mu\in S_n}v_\mu^{(r)}v_{\mu}^{(r)}=1.
 \eea
Interestingly, for large NGF the return-time probability $P(t)$ decays in time at different rates depending on the dimension $n$ over which the diffusion dynamics is defined.
In particular, for a large simplicial complex when $N_{[n]}\to \infty$ we can approximate the return-time probability $P(t)$ as 
\bea
P(t)=\int_{\lambda_1}^{\lambda_{N_{[n]}}} d\lambda\  \rho(\lambda)e^{-\lambda (1-c\delta({\lambda,\lambda_1)}) t}.
\eea  
By inserting the scaling of the density of states given by Eq. (\ref{scrho}), we easily obtain
\bea
P(t)\left\{\begin{array}{lcc}
{A}_{[0]}t^{-d_S^{[0]}/2} & \mbox{for} &n=0,\\
{A}_{[n-1]}t^{-d_S^{[n-1]}/2} +{A}_{[n]}t^{-d_S^{[n]}/2} & \mbox{for} &0<n<d,\\
{A}_{[d-1]}t^{-d_S^{[d-1]}/2} & \mbox{for} & n=d.\\
\label{scPt}
\end{array}\right.
\eea
where $A_{[n]}$  are constants.
In Figure $\ref{Return:fig}$ we provide evidence of the different power-law scaling of the return-time probability $P(t)$ for diffusion processes occurring on the simplices of different dimension $n$ of the NGF.
This result shows that  the diffusion dynamics  defined on nodes, links or triangles of the same instance of simplicial complex generated by the model NGF, can display significantly different dynamical properties. This effect is due to the fact that  the process is  affected by the value of a higher-order spectral dimension that increases with $n$.

\section{Discussion}

Simplicial complexes can sustain dynamics defined not only on nodes but also on higher-order simplices. Linear and non-linear processes  such as diffusion and synchronization  can be extended to higher-order thanks to the higher-order Laplacian. Therefore, the investigation of the spectral properties of the higher-order Laplacian is rather crucial to reveal the properties of higher-order dynamical processes on simplicial complexes. In this work we reveal that the higher-order up and down-Laplacian can display a finite spectral dimension by providing a concrete example where this phenomenon is displayed, the simplicial complex model called ``Network Geometry with Flavor". {In particular, we numerically show} that the up-Laplacians have a spectral dimension that increases with their order $n$ and depends also of the other parameters of the model, i.e. the flavor $s$ and the dimension $d$ of the simplicial complex.
Finally, we show how this spectral property of the higher-order up-Laplacian affects the diffusion dynamics defined on the simplices. {Notably,} we show that different spectral dimensions can cause significant effects in the return-time probability of the diffusion process. {These results} provide evidence that the same simplicial complex can sustain diffusion processes with rather distinct dynamical signatures depending on the dimension $n$ of the simplices over which the diffusion dynamics is defined.

\section*{Acknowledgements}
This research utilized Queen Mary's Apocrita HPC facility, supported by QMUL Research-IT. http://doi.org/10.5281/zenodo.438045.
G. B. thanks Ruben Sanchez-Garcia for interesting discussions and for sharing his code to evaluate the high-order Laplacian.
J. J. T. acknowledges the Spanish Ministry for Science and Technology and the
“Agencia Espa\~nola de Investigaci\'on” (AEI) for financial support under
grant FIS2017-84256-P (FEDER funds).

\end{document}